\documentclass[preprint,amsmath]{epl}
\title{
\large \bf
Size dependence of second-order hyperpolarizability of
finite periodic chains under Su-Schrieffer-Heeger model
}
\author{Shidong Jiang\inst{1} \and Minzhong 
Xu\inst{2}\thanks{Author to whom correspondence should be addressed.
Email: mx200@nyu.edu}}

\institute{
\inst{1} Department of Mathematical Sciences, 
New Jersey Institute of Technology, Newark, NJ 07102\\
\inst{2} Department of Chemistry, New York University, New York, NY 10003
        }

\begin{document}
\maketitle

\begin{abstract}
The second hyperpolarizability 
$\gamma_N(-3\omega;\omega,\omega,\omega)$ of $N$ double-bond finite chain of
trans-polyacetylene 
is analyzed using the Su-Schrieffer-Heeger model 
to explain qualitative features of the
size-dependence behavior of $\gamma_N$. 
Our study shows that $\gamma_N/N$ is {\it nonmonotonic}
with $N$ and that the nonmonotonicity is caused by the dominant
contribution of the intraband transition to $\gamma_N$ in polyenes. 
Several important physical effects are 
discussed to reduce quantitative discrepancies between experimental and 
our results.
\end{abstract}

The size saturation behavior of the second-order 
hyperpolarizability 
$\gamma_N[-(\omega_1+\omega_2+\omega_3);\omega_1,\omega_2,\omega_3]$ 
of finite conjugated polymer 
(especially simple polyene chains) has been extensively studied both 
experimentally\cite{bredas,samuel,ledoux,craig,puccetti} 
and
theoretically\cite{agrawal,mcintyre,beratan,shuai,su1,spano,demelo,heflin,
soos,yaron,zhang,mukamel2,mukamel3,lu,kohler1,kohler2} for the past
several decades. A power law on molecular size with variable power exponent
is often used
to describe the magnitude of the off-resonant nonlinear response in
scaling. That is, $\gamma_N \sim N^b$, where $N$ is the 
number of double bonds\cite{samuel,ledoux}. The scaling is expected to saturate
for the large $N$ (thermodynamic limit), i.e., $b=1$ as $N\rightarrow \infty$.
Several measurements on cubic optical nonlinearity in long-chain polyene 
oligomers\cite{samuel,ledoux} have shown the following three main
features of the size dependence: (i) the exponent $b\sim 3-3.5$ for small $N$;
(ii)  $\gamma_N/N$ 
is {\it nonmonotonic} as a function of $N$, that is,
$\gamma_N/N$ increases with $N$ initially, and after having reached a
maximum value, then gradually decreases (about $10\%$) to the saturation value
as $N$ increases;
(iii) the onset of the saturation occurs at $N\sim 90$ by the {\it nonmonotonic}
fitting. We remark here that the nonmonotonicity of $\gamma_N/N$ has been overlooked by
most theoretical computations, and that in Ref. \cite{ledoux} the experimental data
were fitted by a monotonic curve within experimental error and the onset of the saturation occurs at 
$N\sim 60$ by the {\it monotonic} fitting.

Theoretical studies range from simple tight-binding such as H\"{u}ckel or 
Su-Schrieffer-Heeger (SSH) models\cite{agrawal,mcintyre,beratan,shuai,su1,spano}
to Parier-Parr-Pople (PPP)\cite{soos,yaron}, Hubbard\cite{zhang} and 
electron-hole pairs models\cite{mukamel2,mukamel3}. Detailed
quantitative comparisons between experiments 
and theories are quite difficult due to the following several
reasons. First, most theoretical studies have
been limited to planar {\it trans}-polyenes, while the polyene chains
in the solution are usually `disordered' or `worm-like'\cite{ledoux}. 
Second, experiments used series of polymers that, although related to polyenes,
contain a central phenyle ring and incorporate rings that are saturated
aside from a double bond in the ``polyene''\cite{ledoux}. Third, the optical 
gaps are around
2.3 $eV$ rather than typical polyacetylene's 1.8 $eV$ in most theoretical
studies\cite{ledoux}. Hence, even under the same assumption of the planar 
structure of polyene chains, different models coupled with various 
(analytic and/or numeric) approximations
have predicted vastly different 
$\gamma_N$ and saturation behaviors for polyenes
\cite{agrawal,mcintyre,beratan,shuai,su1,spano,demelo,heflin,soos,
yaron,zhang,mukamel2,mukamel3,lu,kohler1,kohler2}. 

In this letter, based on the SSH model and our previous 
work\cite{mxu1,jiang1}, we have derived an exact
expression for the second-order hyperpolarizability of
third-harmonic-generation (THG) $\gamma_N(-3\omega;\omega,\omega,\omega)$ 
for the finite chain of $N$ double bonds. By choosing the typical
parameters for polyenes,
we notice that the exponent $b$ varies from 2 to 5 for small $N$
($N<21$) in our computation. Our result also shows strong
nonmonotonicity of $\gamma_N/N$ versus $N$. These match the
experimental observations (i) and (ii) mentioned above.
Finally, we observe that there are
large quantitative discrepancies between this theoretical study 
and experiments in observation (iii) and the magnitude of $\gamma_N/N$.

Our computation shows
that $\gamma_N$ can be split into two parts: a positive part due to
intraband transitions and a negative part due to interband
transitions. The positive contribution from intraband transitions
dominates the negative contribution from interband 
transitions. A positive $\gamma_N$ in our computation 
is consistent with all existing polyene experiments
\cite{bredas,samuel,ledoux,craig} which show that 
$\gamma_N$ is positive for any $N$. Moreover, we have carefully
treated the so-called unphysical interference effects 
mentioned in the previous calculation\cite{mukamel3}. Our
results show that they can actually be identified as the boundary
effects. Finally, since the zero-frequency limit $\gamma_N(0;0,0,0)/N$
can not be measured directly, most experimentalists first measure the THG
spectrum $\gamma_N(-3\omega;\omega,\omega,\omega)/N$ by choosing the frequency
close to its three-photon resonance\cite{samuel,ledoux} or the
second-harmonic-generation spectrum 
$\gamma_N(-2\omega;\omega,\omega,0)/N$\cite{craig}, then apply the empirical
extrapolation to obtain $\gamma_N(0;0,0,0)/N$.
This empirical extrapolation largely depends on the three-photon-resonance
of polyenes, but it also needs information on the linear absorption of 
solvent\cite{samuel,ledoux}. Here we have derived the exact expressions
for both the static hyperpolarizability ($\omega=0$) and dynamic 
hyperpolarizability THG ($\omega\ne0$). This could provide us some physical
examinations of those empirical extrapolation methods.
Despite the ignorance of Coulomb
interactions\cite{bredas,demelo,heflin,soos,yaron,zhang,mukamel2,
mukamel3} 
and some other important effects (such as the effects of end
groups\cite{puccetti}, the conformational 
disorder\cite{rossi}, the solvent effect\cite{luo1}, the segments or
short conjugation length\cite{kohler1,kohler2}, etc) that must be
considered in the more accurate quantitative calculations on 
one-dimensional(1D) polymers, our results show that 
the non-interacting SSH model (which is only first approximation of
real physical systems) can nevertheless provide a clear understanding of the
qualitative physical pictures of the saturation behavior.

The Hamiltonian for the SSH model \cite{ssh} is given by:
\begin{eqnarray}
H=-\sum_{l,s} \left[ t_0+(-1)^l 2\alpha u\right]
(\hat{C}_{l+1,s}^{\dag}\hat{C}_{l,s}^{}+\hat{C}_{l,s}^{\dag}\hat{C}_{l+1,s})^{}
+2nKu^2,
\label{Hssh}
\end{eqnarray}
where $t_0$ is the transfer integral or hopping between the nearest-neighbor 
sites, $u$ is the dimerization displacement, $\alpha$ is the changing rate of
the hopping, $n$ is the total number of CH monomers, $K$ is the elastic 
constant, and $\hat{C}_{l,s}^{\dag}(\hat{C}_{l,s})$
creates(annihilates) an $\pi$ electron at site $l$ with spin $s$. 
Each site is occupied by one electron. With the lattice constant $a$ and
the definition of gap parameter $\Delta\equiv 4\alpha u$, we have the
eigenenergies:
\begin{eqnarray}
\varepsilon_v (k)= -\varepsilon_c(k)= 
-\sqrt{\left[ 2 t_0 cos(ka) \right]^2+\left[ \Delta sin(ka)
\right]^2},
\label{ek}
\end{eqnarray}
where $\varepsilon_v(k)$ and $\varepsilon_c(k)$ correspond to eigenenergies in
the valence and conduction bands, respectively. 
To avoid Jahn-Teller effects\cite{soos,kivelson}, we work with the
nondegenerate ground states of H\"{u}ckel chain with $n=4m+2$ sites.
Therefore,
for a polyene chain with N($\equiv n/2$) double-bonds,
the pseudo-momentum vector ${\bf k}=0,\pm\pi/Na,\dots,\pm(2N-1)\pi/(4Na)$.

We resolve this size dependence problem via the following two
steps. First, for each $N$ we fix $t_0$, $\alpha$ and $K$
to find the minimum energy of the ground state in Eq.~\ref{Hssh} by
varying $u$. When the minimum energy is achieved, we denote the
corresponding value of $u$ by $u_0$.
Second, under the static lattice configuration determined by $u_0$, 
we obtain an exact expression for $\gamma_N$ for the dimerized
H\"{u}ckel chain with $N$ double bonds. This two-step treatment takes
into account the fact that $\pi$ electrons make a much greater contribution 
to the $\gamma_N$ in polyene
chain than nuclei and $\sigma$ electrons\cite{bredas}. The Peierls
instability in 
the quasi-1D polymer system\cite{heeger88} is also carefully considered
in this method.

By choosing $t_0=2.5 eV$, $\alpha=4.1 eV/\AA$, $K=18.0 eV/\AA^2$ and 
$a=1.22\AA$, we obtain the
energy gap $E_g \sim 1.89 eV$ and $u_0\sim 0.058\AA$ when $N\to\infty$. Both
values are a little bit larger than those in the reported experiments where
$u_0\sim 0.04 \AA$ and $E_g \sim 1.8 eV$\cite{heeger88}. However, 
the above parameters are not unreasonable since non-interacting
models neglect the important Coulomb interactions in the
1D polymer chains\cite{heeger88}. The inclusion of the strong Coulomb 
interactions are generally expected to reduce the dimerized constant $u_0$,
hence it reduces the gap parameter $\Delta$ and the energy gap 
$E_g$\cite{heeger88} as well.

Table.\ref{nuegtab} shows the relationship between $N$, $u_0$, $2\Delta$ and 
$E_g$. We would like to point out that there is no dimerization distortion for the 
extremely short chain with $N<5$. A Peierls-type distortion
toward a bond-alternated geometric structure with 14 or more $\pi$
electrons has also been reported in the semiempirical calculations of
cyanine dyes\cite{bredas}. Since the dimerization distortion only
exists for a reasonable size of polyene chains, we apply our
calculation of $\gamma_N$ only for $N\ge 5$.

Following the same procedures that have been developed in our previous
work\cite{mxu1,jiang1} and by using the discrete summation of wave 
vectors ${\bf k}$ instead of integrals, we obtain the following
expression for the THG hyperpolarizability for a single finite chain:
\begin{eqnarray}
\label{eq:thg}
&&\gamma_N(-3\omega;\omega,\omega,\omega)
=\displaystyle\frac{e^4a^4}{512\delta^7t_0^3}\sum_{k(occ)}
\left\{\left[\frac{1}{2x^9(x^2-z^2)} -\frac {9}{2x^9( x^2-(3z)^2)}\right]
\right.\\\nonumber
&&-\left. (1-\delta^{2}x^2)(x^2-1) \left[\displaystyle
\frac{216}{x^9(x^2-(3z)^2)}-\frac{252}{x^7(x^2-(3z)^2)^2}
-\displaystyle\frac{24}{x^9(x^2-z^2)}
-\frac{12}{x^7(x^2-z^2)^2}\right]\right\},
\end{eqnarray}
where $\delta\equiv\Delta/2t_0$, $x\equiv\varepsilon_c(k)/\Delta$ and 
$z\equiv\hbar\omega/2\Delta$.
Eq.~\ref{eq:thg}$/(2Na)$ recovers $\chi^{(3)}(-3\omega;\omega,\omega,\omega)$ 
for 
the infinite chain as $N\to\infty$\cite{jiang1}. The first two terms in 
Eq.~\ref{eq:thg} correspond to the contributions from the interband 
transitions while the remaining terms correspond to those from the intraband
transitions. 

For the finite chain, the expressions of $\chi^{(3)}$ or
$\gamma_N/N$
also have another boundary term\cite{jiang2}.  We have neglected this
term in our study due to the following
physical reasons. First, the boundary term disappears when $N\to\infty$.
Second, it causes the so-called
strong unphysical interference effects if it is included\cite{mukamel3}.
Third, the actual measurement of polyene materials in the solvent
solution\cite{samuel,ledoux} are on chain but not on ring structures,
and the random phase on two open ending groups should be expected.
Therefore, the boundary term should not have any observable effect or play
a strong role in real physical environments.

\begin{figure}
\vskip -18pt
\begin{center}
\scalebox{0.5}{\onefigure{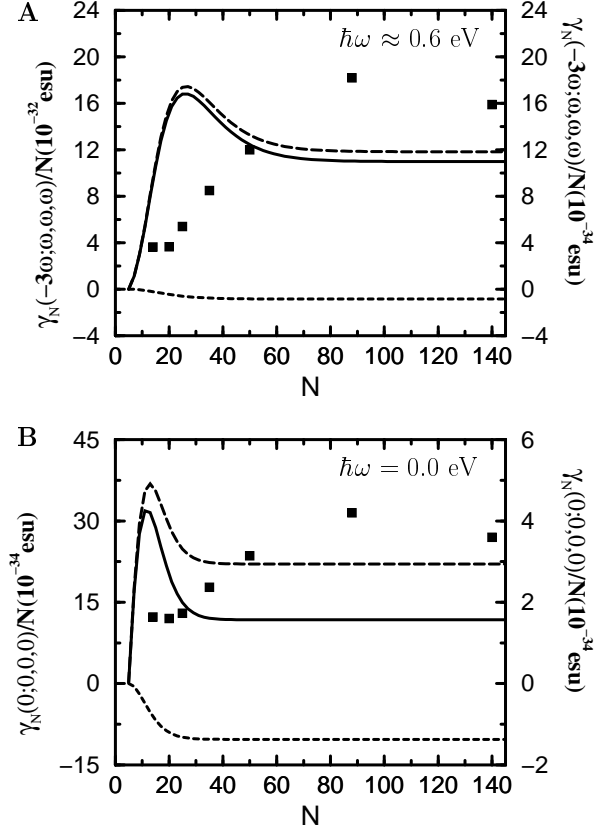}}
\end{center}
\vskip -18pt
\caption{{\bf A.} Computed THG $\gamma_N(-3\omega;\omega,\omega,\omega)/N$ at
$\hbar\omega=0.6 eV$ (left Y scale in units of $10^{-32}$ esu): total
contribution (solid line), intraband contribution (long dashed line) and
interband contribution (short dashed line) versus the experimental THG result
$\gamma_N(3\omega)/N$ (Ref.3) at $\lambda=1.91\mu$m (solid square and the
right Y scale in units of $10^{-34}$esu);
{\bf B.} Computed static $\gamma_N(0;0,0,0)/N$ (left Y scale
in units of $10^{-34}$ esu): total contribution (solid line), intraband
contribution (long dashed line) and interband contribution (short dashed line)
versus the experimental extrapolation static $\gamma_N(0)/N$ (Ref.3) (
solid square and right Y scale in units of $10^{-34}$esu).}
\label{gr:gamma}
\end{figure}

The existing optical experiments are measured under the wavelength 
$\lambda=1.91\mu$m (or $\hbar\omega\sim 0.65
eV$)\cite{samuel,ledoux}. This wavelength is
very close to the edge of three-photon resonance. In order to guarantee the THG 
transition in the off-resonant region,
we have chosen $\hbar\omega=0.6 eV$ in our calculation. 
Substituting the parameters in
Table. \ref{nuegtab} into Eq.~\ref{eq:thg}, we have computed 
$\gamma_N(-3\omega;\omega,\omega,\omega)/N$ for 
$\hbar\omega\equiv 0.6 eV$ and  $\hbar\omega\equiv 0 eV$. The results
are shown in Fig.\ref{gr:gamma}A and in Fig.\ref{gr:gamma}B,
respectively. On the same graphs, we
have also compared them with the existing experimental results\cite{ledoux}.
The graphs clearly show that the positive contribution from intraband
transitions always dominates the negative contribution from interband
transitions for any $N$. The fact that $\gamma_N$ is positive in our
calculation agrees with the reported experiments\cite{bredas}. 
It is also consistent with the theory presented by Agrawal
{\it et. al.}\cite{agrawal}, but differs from 
McIntyre {\it et. al.}'s\cite{mcintyre} and
Beratan {\it et. al.}'s \cite{beratan} results where only the negative 
interband contributions are considered.

Fig.\ref{gr:gamma} shows $\gamma_N/N$ firstly increases until $N$ reaches about
20-30 and then decreases thereafter. Though this trend is parallel to the
{\it nonmonotonic} feature in existing experiments\cite{samuel,ledoux},
it has never been reported in the existing
theories\cite{agrawal,mcintyre,beratan,shuai,su1,spano,demelo,heflin,soos,
yaron,zhang,mukamel2,mukamel3,lu,kohler1,kohler2}.
$\gamma_N/N$ curve has a much longer tail and almost 2 orders larger value for
$\hbar\omega=0.6eV$ than its static limit $\hbar\omega=0.0eV$. It shows that
the three-photon resonance plays a strong role to saturate at a
relatively large $N$.
$\gamma_N(-3\omega;\omega,\omega,\omega)/N$ saturates at $N^\gamma_{sat}\sim 60$ for
$\hbar\omega=0.6eV$ while at $N^\gamma_{sat}\sim 40$ for its
static counterpart. The much larger saturation length and $\gamma_N$ value
at finite freqencies than at the static limit has also been reported in
Luo {\it et. al.}'s work\cite{luo2}.
Our calculated $N^\gamma_{sat}$ mismatches worse at $0eV$ than
at $0.6eV$ with the experiments that report $N^\gamma_{sat}\sim 60$ (without
considering {\it slight nonmonocity}) for both
cases\cite{samuel,ledoux}. The $N^\gamma_{sat}$ difference
between this calculation and experiment for $0eV$ could be understood because
the experiment uses the following extrapolation formula to obtain the
static limit\cite{samuel,ledoux}:
\begin{eqnarray}
\gamma_N(0;0,0,0)=\displaystyle\frac{1-(\lambda_{max}/\lambda)^2}
{1-(3\lambda_{max}/\lambda)^2}\gamma_N(-3\omega;\omega,\omega,\omega),
\label{eq:extrapolate}
\end{eqnarray}
where $\lambda_{max}$ and $\lambda$ correspond to the wavelength of the
maximum of the absorption in solution and the measurement, respectively.
Eq.~\ref{eq:extrapolate} assumes that $\gamma_N(0;0,0,0)$ is largely
dependent on the three-photon process in
$\gamma_N(-3\omega;\omega,\omega,\omega)$. However,
Eq.~\ref{eq:thg} shows that the one-photon process ($(x^2-z^2)^{-1}$ terms)
in $\gamma_N$ plays a much more important role at its static limit,
even if it is very small compared to its three-photon counterpart at
$\hbar\omega=0.6eV$.
The one-photon's contribution was ignored
in the above experimental extrapolation formula.

\begin{figure}
\vskip -10pt
\scalebox{0.4}{\onefigure{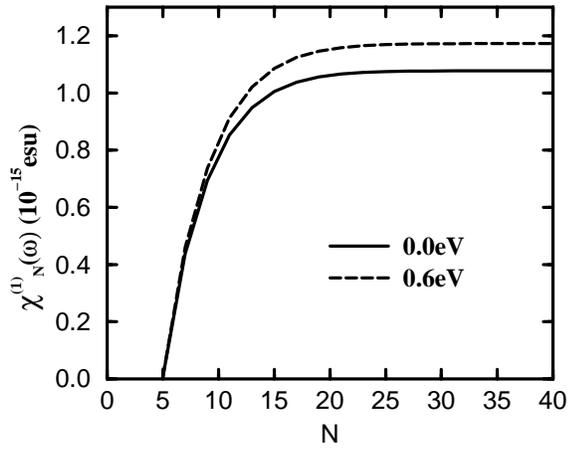}}
\vskip 0pt
\caption{Computed linear susceptibility $\chi^{(1)}(N, \omega)$ versus $N$
at $\hbar\omega=0.0 eV$ and $\hbar\omega=0.6 eV$ for $N\ge5$. }
\label{gr:x1}
\end{figure}

The one-photon's influence
on the saturation behavior in $\gamma_N/N$ is similar to that of linear
susceptibility $\chi^{(1)}_N$. Fig.\ref{gr:x1} is plotted according to
Eq.(3.13) in Ref.\cite{mxu1}. $\chi^{(1)}_N$ saturates at
$N^{\alpha}_{sat}\sim 25\pm 2$. The {\it non-monotonic} $\gamma_N/N$
up-and-down trend is
dominated by the intraband contributions and are closely related with the
linear
saturation $N^{\alpha}_{sat}$. The intraband contributions are
determined by gradient ${\bf k}$ terms or ${\bf k}$-changing rate of
dipole-transitions\cite{agrawal,jiang1}. The intraband $\gamma_N^{intra}$
drastically increases with $N$ until $N$ reaches $N^{\alpha}_{sat}$, then
the increasing of $N$ (corresponding to the additional $\pi$-electron)
lowers the average contribution of $\gamma_N^{intra}$ until
$\gamma_N^{intra}/N$
reaches its static limit. Meanwhile the negative but relatively weak
interband $\gamma_N^{inter}$ increases monotonically with $N$ to its
static limit. This may explain the up-and-down features
of $\gamma_N/N$ in experiments\cite{samuel,ledoux}.

Our theoretical results of $N^{\gamma}_{sat}\sim 40$ and
$N^{\alpha}_{sat}\sim 25$ at $\omega=0$ are in agreement with Lu
{\it et. al.}'s ab initio
calculations of polyene ($N^{\gamma}_{sat}=45\pm5$ and
$N^{\alpha}_{sat}=25\pm2$)\cite{lu}. They are also comparable with the work of
Shuai and Br\'{e}das ($N^{\gamma}_{sat}\sim 50$)\cite{shuai},
Yu and Su ($N^{\gamma}_{sat}\sim 50$)\cite{su1} and
Spano and Soos ($N^{\gamma}_{sat}\sim 30$ for $\delta\approx0.18$)\cite{spano}.
We have applied a factor of $1/2$ on the number of $n$-site carbon to convert
the results in Ref.\cite{shuai,su1,spano}.

\begin{figure}
\vskip -10pt
\scalebox{0.4}{\onefigure{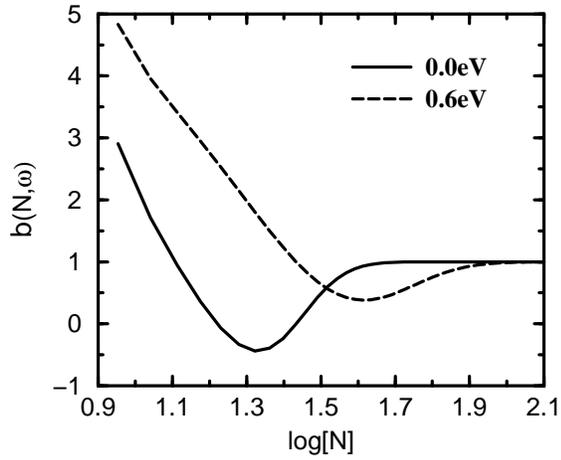}}
\vskip 0pt
\caption{Computed scaling power coefficient
$b(N,\omega)\equiv d[log\gamma_N(-3\omega;\omega,\omega,\omega)]/d[log N]$
versus $log N$ at $\hbar\omega=0.0 eV$ (solid line) and
$\hbar\omega=0.6 eV$ (long dashed line) for $N\ge9$.}
\label{gr:scale}
\end{figure}

Fig.\ref{gr:scale} shows the scaling power $b$ versus $N$. $b$ varies with
both $N$ and $\omega$. It is also very sensitive to some other factors
such as the hopping $t_0$, gap parameter $\Delta$, etc\cite{shuai,su1,spano}.
The scaling power law has also been extensively studied in many theoretical
works\cite{beratan,shuai,su1,spano,demelo,heflin,soos,yaron,zhang,
mukamel2,mukamel3,lu}. Due to the fact that $b$ is a derivative from
$\gamma_N$, here we only make a simple discussion.
For $N$=9, we have $b=4.83$ at $0.6eV$ and $b=2.91$ at the static limit.
The steadily decreasing trend in the graph is quite similar to
Fig.4 ($\delta=0.18$) in Ref.\cite{spano} when $N\le 30$.

Some quantitative discrepancies between this calculation and the
existing experiments\cite{samuel,ledoux} are evident.
Although a much large
or upper bound limit of $\gamma_N$ value should be expected under the
H\"{u}ckel model due to the
ignorance of Coulomb interactions\cite{spano,soos}, theoretical values of
$\gamma_N/N$ are much larger than experimental data 
especially when $\hbar \omega =0.6eV$.
There are many factors to influence this result. One chief factor could be
the conformational behavior in the solvents\cite{ledoux,rossi,luo1}
and the fact that the polyene chain is no longer oriented in
one direction. Hence the worm-like polymer chains may significantly reduce
$\gamma_N$ value by averaging the contribution in $3D$ space. The segment
or short conjugation length treatment of polyene chains may also reduce
$\gamma_N$\cite{kohler1,kohler2}. Another
important factor is the damping or the lifetime of excited states\cite{shuai}.
This may smear off the resonant peak and hence reduce $\gamma_N$.
The optical gap $\Delta$ of experimental ``polyene'' is around 2.3
$eV$ while ours is 1.9 $eV$. Since $\gamma_N/N\sim\Delta^{-6}$\cite{agrawal},
theoretical values would be about $(2.3/1.9)^6\approx3.2$ times as large as the
experimental data due to this factor alone.
Finally, the nondegenerate ground state resulting from the end
groups or fixed ring structures\cite{ledoux,craig,puccetti} in experiments may
also play a role.

In conclusion, the study of the size dependence of $\gamma_N$ based
on the SSH model provides us a solid physical background to
understand the saturation behaviors in polyene systems. Most valuable
features of in the experiments of $\gamma_N$
could be qualitatively explained under the schema of single-electron models.
However, the quantitative comparison with the existing experiments in 1D
polymer system still shows that the SSH model is only first
approximation for real physical systems. Further studies require the
refinement of the model with the consideration of many other important
factors such as the Coulomb interactions, conformational behaviors, segments
effects of chains, solvent effects or the damping factor, etc.
\bibliography{}
\pagebreak

\begin{table}
\caption{
Relationship between number of double bonds $N$, dimerized constant
$u_0$, gap parameter $2\Delta$ and actual energy gap $E_g$.
($t_0=2.5 eV$, $\alpha=4.1 eV/\AA$ and $K=18.0 eV/\AA^2$).
}
\label{nuegtab}
\begin{tabular}{cccc}
$N$ & $u_0(\AA)$ & $2\Delta(eV)$ & $E_g(eV)$ \\ \hline
 5& $0.0073$ & 0.238 & 3.098\\
 7& $0.0481$ & 1.576 & 2.704\\
 9& $0.0542$ & 1.779 & 2.467\\
11& $0.0562$ & 1.843 & 2.314\\
13& $0.0569$ & 1.868 & 2.211\\
15& $0.0572$ & 1.877 & 2.140\\
17& $0.0574$ & 1.882 & 2.088\\
19& $0.0574$ & 1.883 & 2.051\\
21& $0.0574$ & 1.884 & 2.022\\
23& $0.0575$ & 1.885 & 2.000\\
41& $0.0575$ & 1.885 & 1.922\\
81& $0.0575$ & 1.885 & 1.894\\
251& $0.0575$ & 1.885 & 1.886\\
\end{tabular}
\end{table}

\end{document}